# Underdetermined Blind Source Separation for Sparse Signals based on the Law of Large Numbers and Minimum Intersection Angle Rule

XU Peng-fei*, JIA Yin-jie, WANG Zhi-jian

College of Computer and Information, Hohai University, Nanjing 211100, China

**Abstract**：Underdetermined Blind Source Separation(UBSS) is an important issue, for sparse signals, a novel two-step approach for UBSS based on the law of large numbers and minimum intersection angle rule (LM method) is presented. In the first step, the estimation of the mixed matrix is obtained by using the law of large numbers, and the number of source signals is displayed graphically. In the second step, the method of estimating the source signal with the minimum intersection angle rule is proposed. Finally, two simulation results that illustrate the effectiveness of the theoretical results are presented. It has simple principle and good transplantation capability and can be widely applied in various fields of digital signal processing.

**Key words**：Underdetermined Blind Source Separation；Sparse Signal; Law of Large Numbers；Minimum Intersection Angle Rule

## 1. INTRODUCTION

Blind Source Separation (BSS) is a research hotspot in the field of signal processing because it aims to separate unknown source signals from observed mixtures through an unknown transmission channel. The technique of BSS is applied widely in many fields, such as speech signal processing, image processing, radar signal processing, communication systems, data mining, and so on[1,2,3]. The Underdetermined Blind Source Separation (UBSS) is one case of BSS that the number of observed signals is less than the number of source signals. Currently, many researchers have put forward several improved algorithms related to this issue. Van[4] aims at inverting the different nonlinearities, thus reducing the problem to linear underdetermined BSS. To this end, first a spectral clustering technique is applied that clusters the mixture samples into different sets corresponding to the different sources. Kim[5] proposed a novel algorithm based on a single source detection algorithm, which detects time-frequency regions of single-source-occupancy. Xie[6] presented a new time-frequency (TF) underdetermined blind source separation approach based on Wigner-Ville distribution (WVD) and Khatri-Rao product to separate N non-stationary sources from M(M <N) mixtures. Liu[7] exploits the source temporal structure and propose a linear source recovery solution for the UBSS problem which does not require the source signals to be sparse. Zhen[8] uses an effective approach to discover some 1-D subspaces from the set consisting of all the time-frequency (TF) representation vectors of observed mixture signals. Based on the sparse reconstruction model, a single layer perception artificial neural network is introduced into the proposed algorithm introduces, and the optimal learning factor is calculated, which improves the precision of recovery[9]. In this paper, we researched and analyzed the Underdetermined Blind Source Separation for Sparse Signals based on the Law of Large Numbers and Minimum Intersection Angle Rule. The paper is organized as follows. In Section 2, we introduce the UBSS model. A new two-step approach for UBSS is introduced and deduced in Section 3. In Section 4, the simulation experiment that indicates the effectiveness of this method is presented. The final section is a summary of the content of this paper and some questions that need further study.

## 2. UBSS MODEL

The linear, noiseless and memoryless models of blind source separation can be described by formula $x(t) = A * s(t)$.

$$\begin{pmatrix} x_1(t) \\ x_2(t) \\ \vdots \\ x_m(t) \end{pmatrix} = \begin{pmatrix} a_{1,1} \\ a_{2,1} \\ \vdots \\ a_{m,1} \end{pmatrix} s_1(t) + \begin{pmatrix} a_{1,2} \\ a_{2,2} \\ \vdots \\ a_{m,2} \end{pmatrix} s_2(t) + \ldots \begin{pmatrix} a_{1,n} \\ a_{2,n} \\ \vdots \\ a_{m,n} \end{pmatrix} s_n(t) \quad (1)$$

In time $t$, $s(t) = [s_1(t), s_2(t), \cdots, s_n(t)]^T$ represents the $n$ dimensional source signal vector,

$x(t) = [x_1(t), x_2(t), \cdots, x_m(t)]^T$ represents the $m$ dimensional mixed signal vector. The mixed matrix $A = (a_1, a_2, \cdots, a_n)$, $a_i = [a_{1,i}, a_{2,i}, \cdots, a_{m,i}]^T$ is the $i$-th column vector of $A$. When $M > N$, the number of mixed signals is more than that of source signals, which is called overdetermined blind source separation. When $M = N$, the number of mixed signals is equal to the source signal, which is called well posed blind source separation. When $M < N$, the number of mixed signals is less than the source signal, which is called underdetermined blind source separation(UBSS), UBSS is a more realistic and challenging problem. When the problem becomes underdetermined, the mixed system is no longer invertible, so the estimation of the source signal can not be obtained simply by inverting the mixed matrix. The formula (1) can be considered as an underdetermined mixture (see FIG. 1).

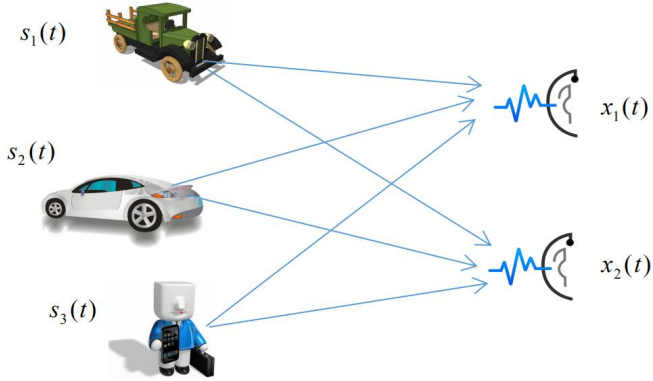

FIG. 1. Underdetermined mixture

UBSS for sparse signals is discussed in this paper. If the signal is sparse, the value of most of the time is zero or close to zero. The sparsity of signals brings great convenience to signal processing. So far, the two-step method [10] is a common method to solve the blind separation of sparse signals, that is, the mixed matrix $A$ is first estimated, and then the source signal is estimated by the reconstruction algorithm. In this paper, a new two-step underdetermined blind source separation method based on the law of Large numbers and Minimum angle rule (LM method) is proposed. In the first step of estimating matrix estimation stage, cluster calculation is performed on uniform sample of the mixed data point by using the Law of Large Numbers, thus one engineering model for estimating the mixing matrix has been established, which can directly estimate the number of source signal. In the second step, the source signal is estimated by way of the minimum intersection angle rule for better separation performance.

## 3. LM METHOD

In general, the only data which we can analyze is the received mixed data. When the source signal is assumed to be sparse, we can separate the source signal by sparsity. The following is an example.

Suppose there are three sparse source signals $s_1$, $s_2$ and $s_3$, they are combined into two mixed signals $x_1$ and $x_2$ according to the following formula (2).

$$\begin{pmatrix} x_1 \\ x_2 \end{pmatrix} = \begin{pmatrix} a_{11} & a_{12} & a_{13} \\ a_{21} & a_{22} & a_{23} \end{pmatrix} \begin{pmatrix} s_1 \\ s_2 \\ s_3 \end{pmatrix} \quad (2)$$

Because the source signal $s_1$, $s_2$ and $s_3$ are sparse, there must be the following situation in $x_1$ and $x_2$.

In time $t = t_1$, only $s_1$ has value. At this time, $x_1^t = a_{11} s_1^t$, $x_2^t = a_{21} s_2^t$, then $x_2^t / x_1^t = a_{21} / a_{11} = a_1$.
In time $t = t_2$, only $s_2$ has value. At this time, $x_1^t = a_{12} s_1^t$, $x_2^t = a_{22} s_2^t$, then $x_2^t / x_1^t = a_{22} / a_{12} = a_2$.
In time $t = t_3$, only $s_3$ has value. At this time, $x_1^t = a_{13} s_1^t$, $x_2^t = a_{23} s_1^t$, then $x_2^t / x_1^t = a_{23} / a_{13} = a_3$.

As long as the signal is sparse, the value of $a_1$, $a_2$ and $a_3$ exists. When the signal sparsity is relatively large, $a_1$, $a_2$ and $a_3$ are the three biggest frequency of occurrence by using statistical analysis. According to Bernoulli's law of large numbers, frequency is close to probability in a certain sense. Therefore, they constitute an estimation matrix $A' = \begin{pmatrix} 1 & 1 & 1 \\ a_1 & a_2 & a_3 \end{pmatrix}$, which can replace the mixed matrix. These data $a_1$, $a_2$ and $a_3$ can be represented by scatter plots, bar graph and so on. Here they are represented by bar graph (see FIG. 5).

First of all, we use bar graph to estimate mixed matrix $A$ and the number of source signals.

The number of sparse source signals and the parameters of the estimation matrix can be observed from the bar graph, which provides a good basis for the next linear programming.

The number of bars in a bar graph reflects the number of sparse source signals, and the corresponding values are $a_1, a_2$ and $a_3$. When the number of sources is changed, the number of bars varies accordingly. To a certain extent, bar graph is more intuitive and precise to describe the sparsity of mixed data than scatter plot. Here are three cases to illustrate, and then we do experiments on these three situations.

If there is only one source signal at any time in $x_1$ and $x_2$, the estimation matrix of Equation (2) is a column in $A'$ at a certain time. The source signals can be restored normally.

If there are two source signals at any time in $x_1$ and $x_2$, the estimation matrix of Equation (2) are two columns in $A'$ at a certain time. The source signals can be restored normally too.

If there are three source signals at any time in $x_1$ and $x_2$, the estimation matrix of Equation (2) is $A'$ at a certain time.

At these moments, the equation is underdetermined, there are infinitely many solutions to the underdetermined equation. The source signals are not well recovered under normal circumstances.

Secondly, the minimum angle rule is used to extract the source signal. Each mixed signal $x$ is the sum of all vectors $a_i s_i$, it can be described in the following formula (3).

$$x = \sum_{i=1}^{N} a_i s_i \qquad (3)$$

At a certain time $t$, the angle of $x_t$ is $\theta_t = \arctan(x_2^t / x_1^t)$ (see FIG. 2). The three angles corresponding to $a_1, a_2$ and $a_3$ are $\theta_1$, $\theta_2$ and $\theta_3$ respectively. The intersection angle between them and $\theta_t$ are $\theta_{t1}$, $\theta_{t2}$ and $\theta_{t3}$ respectively. $\theta_{t1} = |\theta_t - \theta_1|, \theta_{t2} = |\theta_t - \theta_2|, \theta_{t3} = |\theta_t - \theta_3|$.

The two angles closest to $\theta_t$ is selected as the base vectors of time $t$. For example, at the time $t$, if the inequality (4) is satisfied, the vectors with the minimum intersection angle to the vector $\theta_t$ are $\theta_1$ and $\theta_2$.

$$\theta_{t1} \leq \theta_{t2} \leq \theta_{t3} \qquad (4)$$

Accordingly, $a_1$ and $a_2$ are the base vectors of time $t$, it can be described in the following formula (5).

$$x_t = a_1 s_1 + a_2 s_2 \qquad (5)$$

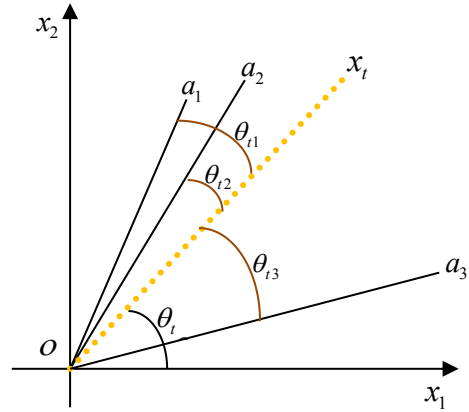

Fig.2 Sketch map of minimum intersection angle

So we decompose the mixed value $x_t$ of time $t$ into a sparse combination of only two source signals and assume that the third source signal has no value at that time. In this way, the optimal solution of equation (2) is obtained, it can be described in the following formula (6).

$$\begin{pmatrix} s_1(t) \\ s_2(t) \end{pmatrix} = A_m^{-1} x = \begin{pmatrix} 1 & 1 \\ a_1 & a_2 \end{pmatrix}^{-1} \begin{pmatrix} x_1(t) \\ x_2(t) \end{pmatrix}$$

$$s_3(t) = 0 \qquad (6)$$

The sampling point of mixed signals is set to $T, t = 1, ..., T$, The mixed signals is decomposed into a combination of several sparse source signals when time $t$ is calculated one by one. The source signals of each time $t$ are obtained by solving equation (6), and then the source signals of the whole observation period are obtained.

In the process of computation, at each time $t$, $A_m$ is a square matrix composed of any two columns of the estimation matrix $A' = \begin{pmatrix} 1 & 1 & 1 \\ a_1 & a_2 & a_3 \end{pmatrix}$. One thing to note here is that the difference between this algorithm and the minimum path algorithm[11] is that the latter selects two vectors closest to it on both sides of $\theta_t$ as the base vectors. However, it is not easy to find the shortest path corresponding to each observation point, which leads to the increase of complexity and the decrease of operation speed. This algorithm avoids this point.

## 4. APPLICATION AND SIMULATION

There are few sparse signals in reality, but ordinary signals can be transformed into sparse signals by various transformations (such as Fourier transform, wavelet transform, etc.). For the convenience of illustration, we experimented with a typical time-domain sparse signal (UWB signal). Ultra-wideband (UWB) is generally based on narrow pulses with extremely low duty cycle, which are sparsely distributed on the time axis. The source signal is generated by the time hopping Ultra-wideband (TH-UWB) system. The number of source signals is set to 3. The pulses used are Gaussian pulse, first-order Gaussian pulse and second-order Gaussian pulse respectively. The pulse width $T_c$=161, the Frame length $T_f = 4T_c$. The number of mixed signals is set to 2, the data length is set to 2898. The exact bit is set to four bits after the decimal point.

The mixed matrix $A$ is generated randomly. Evaluating the performance of blind source separation, a correlation coefficient $C$ is introduced as a performance index[12].

$$C(x,y) = \frac{\text{cov}(x,y)}{\sqrt{\text{cov}(x,x)}\sqrt{\text{cov}(y,y)}} \quad (7)$$

$C(x,y) = 0$ means that x and y are uncorrelated, and the signals correlation increases as $C(x,y)$ approaches unity, the signals become fully correlated as $C(x,y)$ becomes unity.

Here Matlab simulations on above three cases are presented to demonstrate the performance of the LM method for UBSS.

In the first experiment, there are one or two source signals [13] at any time in mixed signals $x_1$ and $x_2$, the mixed matrix $A = \begin{pmatrix} 0.4000 & 0.6000 & 0.3000 \\ 0.8000 & 0.1000 & 0.5000 \end{pmatrix}$, The time domain waveform of the three source signals is shown in Figure 3. The dotted line is used to mark two signals at a certain time. The time domain waveforms of two mixed signals are shown in Figure 4.

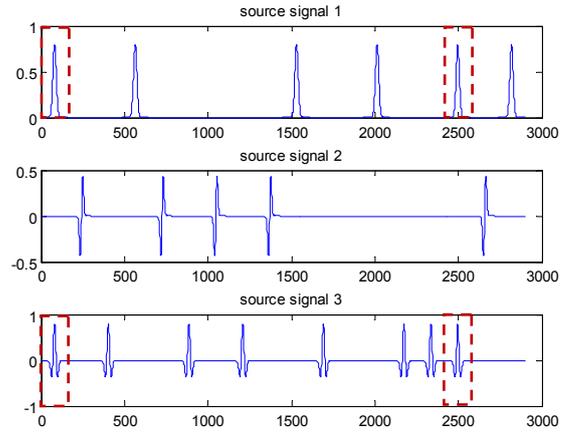

Fig.3 Waveform of three source signals

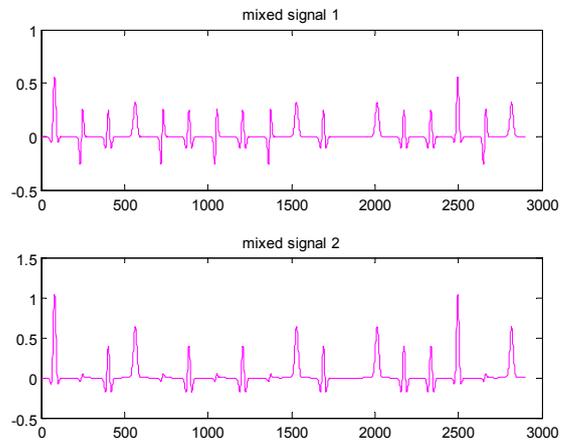

Fig.4 Waveform of two mixed signals

The first step is to find out the estimation matrix, the $a_1$, $a_2$ and $a_3$ in the estimation matrix $A'$ are 0.1667, 1.1667, 2.0000, respectively. The corresponding bar graph is shown in Figure 5.

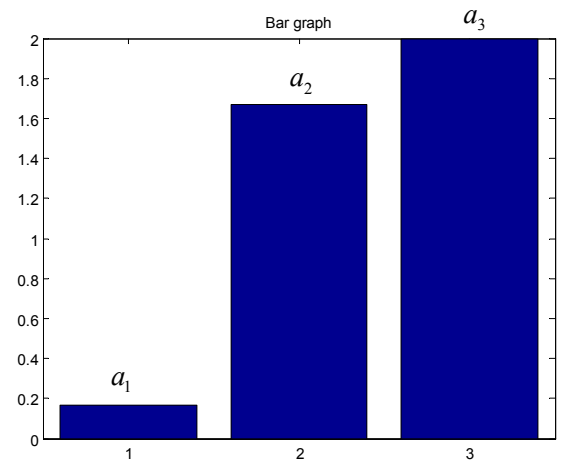

Fig.5    Bar graph

The number of the source signals and the values of the estimation matrix can be intuitively obtained from the diagram.

The second step is to separate the source signal from the mixed signal. The time domain waveforms of the three separated signals are shown in Figure 6.

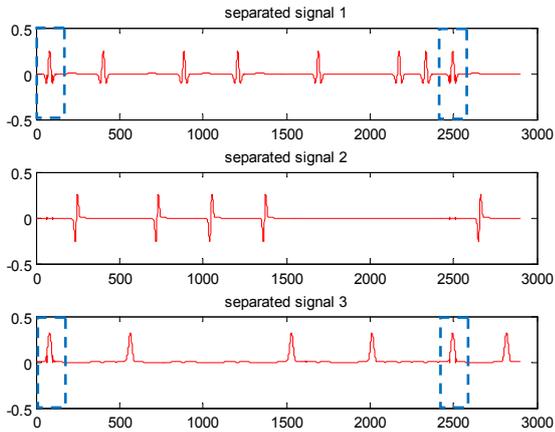

Fig.6 Waveform of three separated signals

As can be seen from Figure 3 and Figure 6, the proposed algorithm can effectively estimate (separate) the source signal from the mixed signal. The correlation coefficients $C$ between them are 0.9996, 0.9946 and 0.9965, respectively. It shows that the separation effect is good.

In the second experiment, there are three source signals at any time in mixed signals $x_1$ and $x_2$. As is shown in the dotted line box. The mixed matrix $A = \begin{pmatrix} 0.5000 & 0.4000 & 0.3000 \\ 0.9000 & 0.2000 & 0.6000 \end{pmatrix}$. The time domain waveform of the three source signals is shown in Figure 7. The time domain waveforms of two mixed signals are shown in Figure 8.

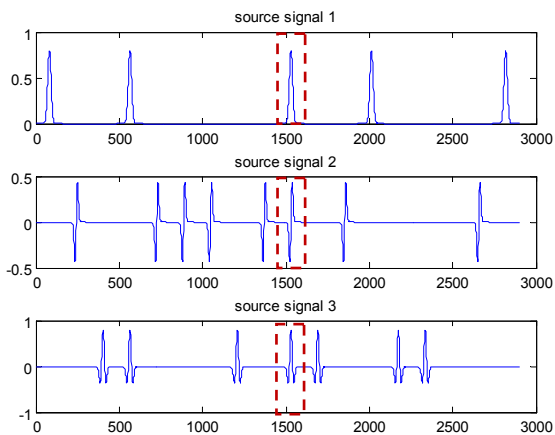

Fig.7 Waveform of three source signals

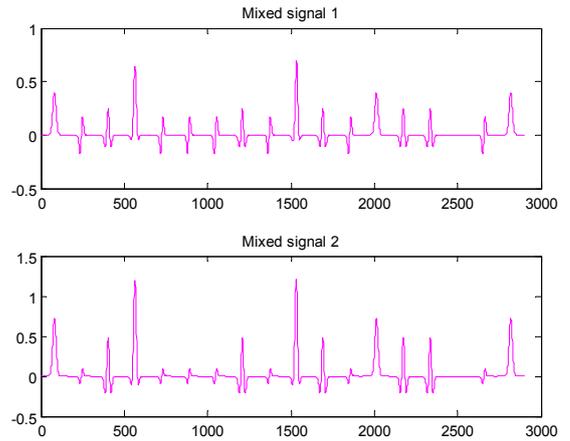

Fig.8 Waveform of two mixed signals

The first step is to find out the estimation matrix, the $a_1$, $a_2$ and $a_3$ in the estimation matrix $A'$ are 0.5000, 2.0000, 1.8000, respectively. The corresponding bar graph is similar to figure 5, which is not repeated here.

The second step is to separate the source signal from the mixed signal. The time domain waveforms of the three separated signals are shown in Figure 9.

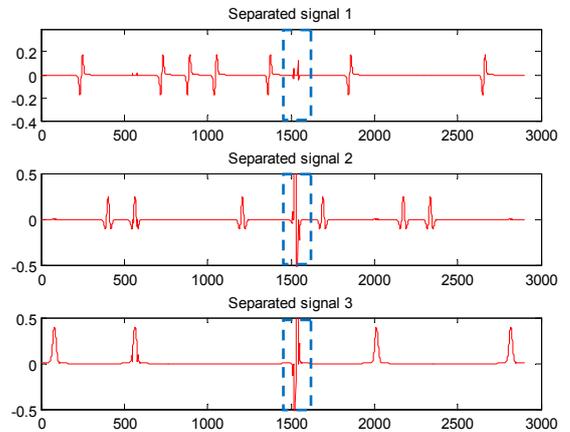

Fig.9 Waveform of three separated signals

As can be seen intuitively from Figure 7 and 9, source waveforms can be separated normally except in three overlapping regions. Although we estimate the mixing matrix, the corresponding waveforms can not be separated at these times.

Two schemes are proposed to solve this problem: (1) Using signal reconstruction technique, the known sequence length is used to "truncate" the sequence in time domain, and the signal is reconstructed from the undistorted part of the data. Signal reconstruction technology is widely used in voice, communication, control and so on.

(2) Each separated signal is regarded as a signal after passing through a certain channel. In both wired and wireless channels, there will be interference or fast fading, which makes the signal at the receiver has error code or distortion. So it can be corrected or deinterleaved to recover this part of the error code or distortion. And then restore the whole signal.

5. CONCLUSION

For UBSS problems, sparse component analysis (SCA) is a good choice. The two-step method is the most representative method of SCA. It includes the estimation of the mixing matrix and the source signal recovery. A novel two-step approach for UBSS based on LM method is presented in this paper. In the first stage, according to the law of large numbers, each data point is clustered and a bar graph is proposed to represent the value of the estimation matrix and the number of source signals. In the second stage, the source signal is estimated by the minimum intersection angle rule. This method is not only suitable for sparse ultra wideband communication signals, but also suitable for other types of sparse signals. Because the source signal is not ideal sparse resulting in local distortion of the separated waveform, if the signal waveform has some regular repetition, signal reconstruction technology can be used to restore the original waveform, or through the channel error correction, de-interleaving technology to restore the source signal. At present, the algorithm is limited to instantaneous mixing, and the case of underdetermined convolutional mixing needs further study and discussion.

REFERENCES

[1] Mirzaei S，Van Hamme H，Norouzi Y.Blind audio source counting and separation of anechoic mixtures using the multichannel complex NMF framework[J].Signal Processing，2015，115：27-37.

[2] C. Deng, Y. Wei, Y. Shen, Q. Su and J. Xu, "An improved approach of blind source separation for delayed sources using taylor series expansion," 2017 3rd IEEE International Conference on Computer and Communications (ICCC), Chengdu, 2017, pp. 284-288.

[3] Tichy O，Smidl V.Bayesian blind separation and deconvolution of dynamic image sequences using sparsity priors[J].IEEE Transactions on Medical Imaging，2015，34（1）：258-266.

[4] S. Van Vaerenbergh and I. Santamaria, "A spectral clustering approach to underdetermined postnonlinear blind source separation of sparse sources," in IEEE Transactions on Neural Networks, vol. 17, no. 3, pp. 811-814, May 2006.

[5] S.Kim and C. D. Yoo, "Underdetermined Blind Source Separation Based on Subspace Representation," in IEEE Transactions on Signal Processing, vol. 57, no. 7, pp. 2604-2614, July 2009.

[6] S. Xie, L. Yang, J. Yang, G. Zhou and Y. Xiang, "Time-Frequency Approach to Underdetermined Blind Source Separation," in IEEE Transactions on Neural Networks and Learning Systems, vol. 23, no. 2, pp. 306-316, Feb. 2012.

[7] B. Liu, V. G. Reju and A. W. H. Khong, "A Linear Source Recovery Method for Underdetermined Mixtures of Uncorrelated AR-Model Signals Without Sparseness," in IEEE Transactions on Signal Processing, vol. 62, no. 19, pp. 4947-4958, Oct.1, 2014.

[8] L. Zhen, D. Peng, Z. Yi, Y. Xiang and P. Chen, "Underdetermined Blind Source Separation Using Sparse Coding," in IEEE Transactions on Neural Networks and Learning Systems, vol. 28, no. 12, pp. 3102-3108, Dec. 2017.

[9] W. Fu, B. Nong, X. Zhou, J. Liu and C. Li, "Source recovery in underdetermined blind source separation based on artificial neural network," in China Communications, vol. 15, no. 1, pp. 140-154, Jan. 2018.

[10] Li Y, Amari A, Cichocki A. Underdetermined blind source separation based on sparse representation[J]. IEEE Trans Signal Process,2006,54(2):423-437.

[11] RBofill，M.zibulevsky. Underdetermined source separation using sparse representation[J]. Signal Proeessing，2001，81,PP.2353-2362.

[12] Peled R, Braun S, Zacksenhouse M. A blind deconvolution separation of multiple sources, with application to bearing diagnostics[J]. Mechanical Systems & Signal Processing, 2005, 19(6):1181-1195.

[13] JIA Yin-jie, XU Peng-fei. Underdetermined Blind Source

Separation applied to the Ultra-wideband communication signals[J]. Journal of Nanjing University of Posts and Telecommunications (Natural Science Edition), 2011, 31(1):23-28.